\documentclass{elsart}
\usepackage{epsf}
\usepackage{epsfig}
\usepackage{graphicx}
\usepackage{dcolumn}
\usepackage{longtable}
\usepackage{amssymb}

\newcommand{\BE}{\begin{equation}}
\newcommand{\EE}{\end{equation}}
\newcommand{\BA}{\begin{eqnarray}}
\newcommand{\EA}{\end{eqnarray}}

\def\be{\begin{equation}}
\def\ee{\end{equation}}
\def\bea{\begin{eqnarray}}
\def\eea{\end{eqnarray}}

\begin{document}

\begin{frontmatter}
\input epsf

\title{{\Large\bf{Inflation Induced Planck-Size Black Hole
Remnants As Dark Matter}}}

\author{Pisin Chen}

\address{Stanford Linear Accelerator Center \\ Stanford University, Stanford, CA 94309, USA
     }

\maketitle

\begin{abstract}
While there exist various candidates, the identification of dark
matter remains unresolved. Recently it was argued that the
generalized uncertainty principle (GUP) may prevent a black hole
from evaporating completely, and as a result there should exist a
Planck-size BHR at the end of its evaporation. We speculate that
the stability of BHR may be further protected by supersymmetry in
the form of extremal black hole. If this is indeed the case and if
a sufficient amount of small black holes can be produced in the
early universe, then the resultant BHRs can be an interesting
candidate for DM. We demonstrate that this is the case in the
hybrid inflation model. By assuming BHR as DM, our notion imposes
a constraint on the hybrid inflation potential. We show that such
a constraint is not fine-tuned. Possible observational signatures
are briefly discussed.
\end{abstract}

\begin{keyword}{primordial black holes : general
--- (cosmology:) dark matter}
\PACS{ 95.35.+d }
\end{keyword}
\end{frontmatter}

\section{Introduction}

It is by now widely accepted that dark matter (DM) constitutes a
substantial fraction of the present critical energy density in the
universe. However, the nature of DM remains an open problem. There
exist many DM candidates, most of them are non-baryonic weakly
interacting massive particles (WIMPs), or WIMP-like
particles\cite{gondolo}. Figure 1 shows the masses and cross
sections of WIMP (or WIMP-like) candidates\cite{roszkowski}. By
far the DM candidates that have been more intensively studied are
the lightest supersymmetric (SUSY) particles such as neutralinos
or gravitinos, and the axions (as well as the axinos). There are
additional particle physics inspired DM candidates\cite{gondolo}.
A candidate which is not as closely related to particle physics is
the relics of primordial black holes
(PBHs)\cite{zeldovich,hawking71}. Earlier it was suggested that
PBHs are a natural candidate for DM\cite{macgibbon}. More recent
studies\cite{carr94} based on the PBH production from the ``blue
spectrum" of inflation demand that the spectral index $n_s\sim
1.3$, but this possibility may be ruled out by recent WMAP
experiment\cite{CMB}. There are also other studies on the idea of
PBH as DM\cite{Khlopov}.

\begin{figure}[th]
\centerline{\psfig{file=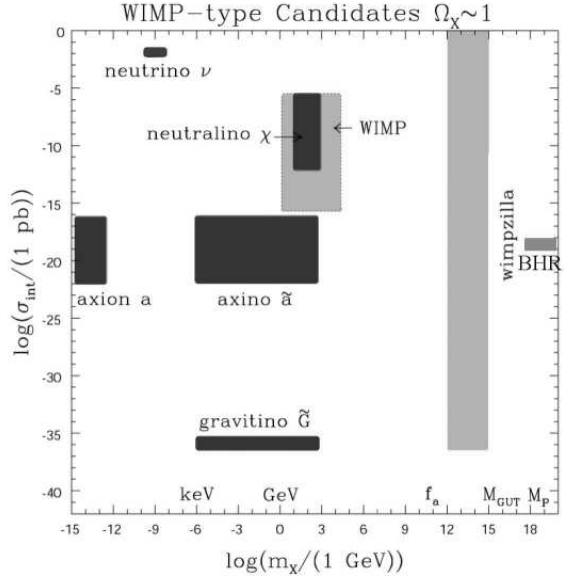,width=4.0in}}
\caption{Cross sections and masses of WIMP dark matter candidates.
Black hole remnant (BHR) is included.}
\end{figure}

One weakness of the notion of PBH as DM is the ambiguity on the
final property of small black holes. The standard view of black
hole thermodynamics\cite{bekenstein1,hawking} does not provide an
answer as to whether a small black hole should evaporate entirely,
or leave something else behind, which we refer to as a black hole
remnant (BHR). Numerous calculations of black hole radiation
properties have been made from different points of
view\cite{wilczek}, and some hint at the existence of remnants,
but none appear to give a definitive answer.

In a recent paper\cite{acs}, a generalized uncertainty principle
(GUP)\cite{mead,veneziano,maggiore,adler} was invoked to argue
that the total evaporation of a black hole may be prevented, and
as a result there should exist a black hole remnant with Planck
mass and size. Here we speculate that the stability of such BHR
may be further protected by supersymmetry, in the form of the
extremal black hole\cite{shmakova}. Such a BHR is totally inert,
with no attributes other than gravitational interaction, and is
thus a natural candidate for DM. It remains unclear, however,
whether such a notion can be smoothly incorporated into the
standard cosmology.

We note that certain inflation models naturally induce a large
number of small black holes. As a specific example, we demonstrate
that the hybrid inflation\cite{linde91,copeland,lyth99} cosmology
can in principle yield the necessary abundance of primordial BHRs
for them to be the primary source of dark matter. We show that
such a construction is not fine-tuned.

\section{Generalized Uncertainty Principle and Black Hole Remnant}

As a result of string theory\cite{veneziano}, or noncommutative
spacetime algebra\cite{maggiore,yoneya}, or general considerations
of quantum mechanics and gravity\cite{mead,adler}, the standard
uncertainty principle must be modified when the gravity effect is
included. A heuristic derivation may be made on dimensional
grounds. Consider a particle such as an electron being observed by
means of a photon with momentum $p$. The usual Heisenberg argument
leads to an electron position uncertainty given by $\hbar/\Delta
p$. But we should add to this a term due to the gravitational
interaction between the electron and the photon, and that term
must be proportional to $G$ times the photon energy, or $Gpc$.
Since the electron momentum uncertainty $\Delta p$ will be of
order of $p$, we see that on dimensional grounds the extra term
must be of order $G\Delta p/c^3$. Note that there is no $\hbar$ in
the extra term when expressed in this way. The effective position
uncertainty is therefore
\begin{equation}
\Delta x\geq \frac{\hbar}{\Delta p}+ \xi^2l_p^2\frac{\Delta
p}{\hbar} \ , \label{eq:B}
\end{equation}
where $l_p=(G\hbar/c^3)^{1/2}\approx 1.6\times 10^{-33}$cm is the
Planck length. Here the dimensionless coefficient $\xi\sim
{\mathcal{O}}(1)$ can be considered either as related to the string
tension from the string theory's motivation, or simply as a factor
to account for the imprecision of our heuristic derivation which
can only be fixed by a precise theory of quantum gravity in the
future. Note that Eq.(\ref{eq:B}) has a minimum value of $\Delta
x_{min}=2\xi l_p$, so the Planck length (or $\xi l_p$) plays the
role of a fundamental length.

The Hawking temperature for a spherically symmetric black hole may
be obtained in a heuristic way with the use of the standard
uncertainty principle and general properties of black
holes\cite{acs}. In the vicinity of the black hole surface there
is an intrinsic uncertainty in the position of any
vacuum-fluctuating particle of about the Schwarzschild radius,
$\Delta x\approx r_s=2GM_{\rm BH}/c^2$, due to the behavior of its
field lines\cite{adler2} - as well as on dimensional grounds. This
leads to a momentum uncertainty $\Delta p$. Identifying $\Delta p
c$ as the characteristic energy of the emitted photon, and thus as
a characteristic temperature (with the insertion of a calibration
factor $1/4\pi$), one arrives at the celebrated Hawking
temperature, $T_{\rm H}=\hbar c^3/8\pi GM_{\rm BH}$.

Applying the same argument, but invoking the GUP, one then finds a
modified black hole temperature,
\begin{equation}
T_{\rm GUP} = \frac{m_pc^2}{4\pi}\frac{\mu}{\xi^2}
\Big[1\mp\sqrt{1-\xi^2/\mu^2}\Big]\ , \label{eq:H}
\end{equation}
where $\mu\equiv M_{BH}/m_p$ is the BH mass in units of the Planck
mass: $m_p=(\hbar c/G)^{1/2}\approx 1.2\times 10^{19}$GeV. This
agrees with the Hawking result for large mass if the negative sign
is chosen, whereas the positive sign has no evident physical
meaning. Note that the temperature becomes complex and unphysical
for mass less than $\xi m_p$ and Schwarzschild radius less than
$2\xi l_p$, the minimum size allowed by the GUP. At $\mu=\xi$,
$T_{\rm GUP}$ is finite but its slope is infinite, which
corresponds to zero heat capacity of the black hole. The BH
evaporation thus comes to a stop, and what remains is a inert BHR
with mass $\mu=\xi$.

If there are $g$ species of relativistic particles, then the BH
evaporation rate (assuming the Stefan-Boltzmann law) is
\begin{equation}
\dot{\mu}=-\frac{16g}{t_{ch}}\frac{\mu^6}{\xi^8}\Big[1-\sqrt{1-\xi^2/\mu^2}
\Big]^4 \ , \label{eq:I}
\end{equation}
where $t_{ch}=60(16)^2\pi t_p\approx 4.8\times 10^4 t_p$ is a
characteristic time for BH evaporation, and $t_p=(\hbar
G/c^5)^{1/2}\approx 0.54\times 10^{-43}$sec is the Planck time.
Note that the energy output given by Eq.(\ref{eq:I}) is finite at
the end point where $\mu=\xi$, i.e.,
$d\mu/dt\vert_{\mu=\xi}=-16g/(\xi^2t_{ch})$. Thus the hole with an
initial mass $\mu_i$ evaporates to a remnant in a time given by
\begin{eqnarray}
\tau&=&\frac{t_{ch}}{16g}\Big[\frac{8}{3}\mu_i^3
+\frac{8}{3}(\mu_i^2-\xi^2)^{3/2}-4\xi^2(\mu_i^2-\xi^2)^{1/2} \cr
&& \quad\quad -8\xi^2\mu_i+4\xi^3\cos^{-1}\frac{\xi}{\mu_i}
+\frac{19}{3}\xi^3-\frac{\xi^4} {\mu_i}\Big] \cr &\approx&
\frac{\mu_i^3}{3g}t_{ch}, \quad\quad \mu_i\gg 1\ . \label{eq:J}
\end{eqnarray}
The evaporation time in the $\mu_i\gg 1$ limit agrees with the
standard Hawking picture.

\section{The Issue of Black Hole Remnant Stability}

Even if the GUP may prevent a small black hole from total
evaporation, it remains unclear whether the resultant BHR will be
prohibited from decaying into the vacuum. In our previous
work\cite{acs} we argued that the total collapse of a black hole
may be prevented by dynamics (i.e., the GUP) and not by symmetry,
just like the prevention of hydrogen atom from collapse by the
standard uncertainty principle\cite{shankar}. In a closer look the
hydrogen atom argument may be only fortuitous, as there exist
counter-examples such as the finite lifetime of the positronium.
Therefore to protect the stability of the BHR, the existence of a
symmetry principle in the system appears essential.

In this regard supersymmetry, in particular supergravity, stands a
very good chance of providing such a protection to BHR. It is
well-known that the no-hair theorem allows for only three
attributes of a classical black hole, namely, its mass, charge and
angular momentum. For the extreme Kerr-Newman BH, we have the
limiting case where
\begin{equation}
M^2=Q^2+a^2\ , \label{eq:K1}
\end{equation}
where $M$ is the BH mass, $Q$ the BH charge associated with
certain gauge symmetry, and $a\equiv S/M$ is the BH angular
momentum per unit mass. In the special case where the BH has no
angular momentum, it reaches the extremal condition of $M=Q$. It
has been shown that in certain specific realizations of
supergravity, there exist extremal black hole
solutions\cite{shmakova}. As supergravity ``charge" is in units of
Planck mass, this condition dictates that the BH mass is bounded
by the Planck mass. We speculate that the extremal condition
should remain intact even when SUSY is broken.

If the primordial black holes were generated in the very early
epoch of the universe, such as the one immediately following
inflation, it is likely that SUSY was still unbroken. Then the
PBHs so generated could be either straight-forwardly the SUSY
extremal BHs governed by supergravity, or the small but classical
BHs described above. We believe that, governed by the GUP, the
latter would soon reduce to Planck-size BHRs whose final state
then coincides with the extremal BH condition. In either case we
expect the existence of BHRs at Planck size. It is unclear,
however, whether this notion can be physically realized. More
efforts are evidently needed beyond the simple-minded comments
made here. A study in this direction based on string theory is
currently underway\cite{CDS}.

\section{Hybrid Inflation and Black Hole Production}

We now return to the scenario of a semi-classical BH and combine
this notion with the hybrid inflation proposed by A.
Linde\cite{linde91,bellido}. In the hybrid inflation model two
inflaton fields, $(\phi,\psi)$, are invoked. Governed by the
inflation potential,
\begin{equation}
V(\phi,\psi)=\Big(M^2-\frac{\sqrt{\lambda}}{2}\psi^2\Big)^2
+\frac{1}{2}m^2\phi^2+\frac{1}{2}\gamma\phi^2\psi^2\ ,
\label{eq:K2}
\end{equation}
$\phi$ first executes a ``slow-roll" down the potential, and is
responsible for the more than 60 e-folds expansion while $\psi$
remains zero. When $\phi$ eventually reduces to a critical value,
$\phi_c=(2\sqrt{\lambda}M^2/\gamma)^{1/2}$, it triggers a phase
transition that results in a ``rapid-fall" of the energy density
of the $\psi$ field, which lasts only for a few e-folds, that ends
the inflation.

 The equations of motion for the fields are
\begin{eqnarray}
\ddot{\phi}+3H\dot{\phi}&=&-(m^2+\gamma\psi^2)\phi\ , \label{eq:K}
\cr \ddot{\psi}+3H\dot{\psi}&=&(2\sqrt{\lambda}M^2-\gamma\phi^2-
\lambda\psi^2)\psi\ , \label{eq:L}
\end{eqnarray}
subject to the Friedmann constraint,
\begin{equation}
H^2=\frac{8\pi}{3m_p^2}\Big[V(\phi,\psi)+\frac{1}{2}\dot{\phi}^2+
\frac{1}{2}\dot{\psi}^2\Big]\ . \label{eq:M}
\end{equation}
The solution for the $\psi$ field in the small $\phi$ regime,
measured backward from the end of inflation, is
\begin{equation}
 \psi(N(t))= \psi_e\exp(-sN(t)) \ , \label{eq:N}
\end{equation}
where $N(t)=H_*(t_e-t)$ is the number of e-folds from $t$ to $t_e$
and $s=-3/2+(9/4+2\sqrt{\lambda}M^2/H_*^2)^{1/2}$ and
$H_*\simeq\sqrt{8\pi/3}M^2/m_p$.

We now show how a large number of small black holes can result
from the second stage of inflation. Quantum fluctuations of $\psi$
induce variations of the starting time of the second stage
inflation, i.e., $\delta t = \delta\psi/\dot{\psi}$. This
translates into perturbations on the number of e-folds, $\delta
N=H_*\delta\psi/\dot{\psi}$, and therefore the curvature
contrasts.

It can be shown that\cite{liddle} the density contrast at the time
when the curvature perturbations re-enter the horizon is related
to $\delta N$ by
\begin{equation}
\delta\equiv \frac{\delta\rho}{\rho}=\frac{2+2w}{5+3w}\delta N \ ,
\label{eq:O}
\end{equation}
where $p=w\rho$ is the equation of state of the universe at
reentry. From Eq.(\ref{eq:N}), it is easy to see that
$\dot{\psi}=sH_*\psi$. At horizon crossing, $\delta\psi\sim
H_*/2\pi$. So with the initial condition (at $\phi=\phi_c$)
$\psi\sim H_*/2\pi$, we find that $\delta N\sim 1/s$. Thus
\begin{equation}
\delta\sim \frac{2+2w}{5+3w}\frac{1}{s}\ . \label{P}
\end{equation}
As $w$ is always of order unity, we see that the density
perturbation can be sizable if $s$ is also of order unity. With an
initial density contrast $\delta(m)\equiv \delta\rho/\rho\vert_m$,
the probability that a region of mass $m$ becomes a PBH
is\cite{carr75}
\begin{equation}
P(m) \sim \delta(m)e^{-w^2/2\delta^2} \ . \label{eq:Q}
\end{equation}

Let us assume that the universe had inflated $e^{N_c}$ times
during the second stage of inflation. From the above discussion we
find\cite{bellido}
\begin{equation}
e^{N_c}\sim \Big(\frac{2m_p}{sH_*}\Big)^{1/s}\ . \label{eq:R}
\end{equation}
At the end of inflation the physical scale that left the horizon
during the phase transition is $H_*^{-1}e^{N_c}$. If the second
stage of inflation is short, i.e., $N_c\sim {\mathcal O}(1)$, then
the energy soon after inflation may still be dominated by the
oscillations of $\psi$ with $p=0$, and the scale factor of the
universe after inflation would grow as $(tH_*)^{2/3}$. The scale
$(tH_*)^{2/3}H_*^{-1}e^{N_c}$ became comparable to the particle
horizon ($\sim t$), or $t\sim (tH_*)^{2/3}H_*^{-1}e^{N_c}$, when
\begin{equation}
t\sim t_h=H_*^{-1}e^{3N_c} \ . \label{eq:S}
\end{equation}
At this time if the density contrast was $\delta \sim 1$, then BHs
with size $r_s\sim H_*^{-1}e^{3N_c}$ would form with an initial
mass
\begin{equation}
\mu_i\sim \frac{m_p}{H_*}e^{3N_c}\equiv \alpha
\frac{m_p}{H_*}\Big(\frac{2m_p}{sH_*}\Big)^{3/s} \ . \label{eq:T}
\end{equation}
A dimensionless parameter $\alpha$ is introduced to account for
the dynamic range of the gravitational collapse. Since $H_*$
depends on $M$ while $s$ on $M$ and $\lambda$, the initial BH mass
depends only on the mass and the coupling in the $\psi$-sector of
the hybrid inflation.

\section{Black Hole Remnants as Dark Matter}

By identifying BHRs as DM and assuming hybrid inflation as the
progenitor of PBHs, we in effect impose a constraint between $H_*$
and $s$ (or equivalently, $M$ and $\lambda$). Though constrained,
these parameters are not so fine-tuned, as we will show in the
following analysis.

We wish to estimate the present abundance of the BHRs created by
hybrid inflation. To do so we should track the evolution of the
post-inflation PBHs through different cosmological epochs. The
newly introduced ``black hole epoch" ($t_h \leq t \leq \tau$)
would in principle involve evaporation and mergers of PBHs as well
as their accretion of radiation; the details of which can be
intricate. For the purpose of a rough estimate and as a good
approximation, however, one is safe to neglect these detailed
dynamics and only keep track of the BH evaporation throughout the
BH epoch. We assume that the universe was matter dominant at the
time $t_h$ when PBHs were formed. Due to Hawking evaporation, the
universe would later become radiation dominant. Since the rate of
BH evaporation rises sharply in its late stage, the crossing time
$t_x$ of this transition can be roughly estimated by integrating
Eq.(3) from $\mu_i$ to $\mu_i/2$. This gives $t_x\sim 7/8\tau$.
When the Hubble expansion effect is included, where the radiation
density dilutes faster than the matter density by one power of the
scale factor, the crossing time would be even closer to $\tau$.
For our purpose, we can simply assume that the entire BH epoch was
matter dominant.

The radiation to matter density ratio at the end of BH epoch, with
Hubble expansion included, can then be estimated as
\begin{equation}
\frac{\Omega_{\gamma,\tau}}{\Omega_{{\rm BHR},\tau}}\sim
\frac{1}{\xi}\int_{t_h}^{\tau}dt\dot{\mu}\Big(\frac{t}{\tau}\Big)^{2/3}
\equiv \beta\frac{\mu_i}{\xi} \ .  \label{eq:V}
\end{equation}
Here we introduce another parameter $\beta$ to account for a range
of possible minor variations of the evolution due to the dynamical
details in the black hole epoch. Not surprisingly, the density
ratio is just roughly the initial BH mass over what remains in its
remnant. As we will see below, the typical BH mass in our
scenario, while small in astrophysical sense, is nonetheless much
larger than the Planck mass, i.e., $\mu_i\gg 1$. Furthermore, the
effective reheating temperature through Hawking evaporation in
this case would be much higher than the energy scales associated
with the standard model of particle physics and baryogenesis. We
thus assume that the standard cosmology would resume after the
black hole epoch. To conform with the standard cosmology, our
assumption that DM is predominantly contributed from BHR demands
that, by the time $t \sim t_{eq} \sim 10^{12}$sec the density
contributions from radiation and BHR should be about equal,
namely,
\begin{equation}
\frac{\Omega_{\gamma,t_{eq}}}{\Omega_{{\rm BHR},t_{eq}}}\sim 1\sim
\Big(\frac{\tau}{t_{eq}}\Big)^{1/2}\frac{\Omega_{\gamma,\tau}}{\Omega_{{\rm
BHR},\tau}}=
\Big(\frac{\tau}{t_{eq}}\Big)^{1/2}\beta\frac{\mu_i}{\xi} \ .
\label{eq:W}
\end{equation}
Since $\tau$ is uniquely determined by the initial BH mass $\mu_i$
(cf. Eq.(\ref{eq:J})), the above condition translates into a
constraint on $H_*$ and $s$ in hybrid inflation through
Eq.(\ref{eq:T}):
\begin{equation}
\frac{m_p}{H_*}\Big(\frac{2m_p}{sH_*}\Big)^{3/s}\sim
\Big(\frac{3g\xi^2}{\alpha^5\beta^2}\frac{t_{eq}}{t_{ch}}\Big)^{1/5}\
. \label{eq:X}
\end{equation}
Figure ~\ref{BHR as DM Parameter Space} shows the region in the
$(H_*,s)$ parameter space that satisfies the above condition. We
assume $g=100$ and $\xi=1$, and let $0.3\leq\alpha\leq 3$ and
$0.3\leq\beta\leq 3$. The allowed $(H_*,s)$ values are shown in
the darkened region. We see that within the constraint that $s$ be
of the order unity so that the metric perturbation $\delta$ be not
exponentially small, there exists a wide range of $H_*$ that could
produce the right amount of PBHs.

\begin{figure}[th]
\centerline{\psfig{file=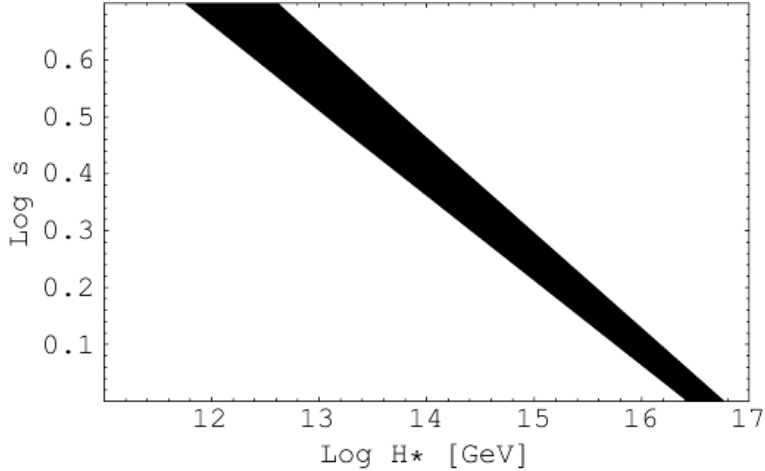,width=4.0in}}
\vspace*{8pt} \caption{Region in the hybrid inflation $(H_*,s)$
parameter space where the induced black hole remnants would
provide the right abundance for dark matter.}\label{BHR as DM
Parameter Space}
\end{figure}

As an example, we take $H_*\sim 5\times 10^{13}$ GeV and $s\sim
3$. Assume that the universe was matter-dominated when the
curvature perturbation reentered the horizon. Then the density
contrast is $\delta \sim 1/7$, and the fraction of matter in the
BH is $P(m)\sim 10^{-2}$. From Eq.(\ref{eq:R}), $e^{N_c}\sim 54$.
So the total number of e-folds is $N_c\sim 4$. The black holes
were produced at the moment $t_h\sim 2\times 10^{-33}$ sec, and
had a typical mass $M_{{\rm BH}i}\sim 4\times 10^{10}m_p$. Let
$g\sim 100$. Then the time it took for the BHs to reduce to
remnants, according to Eq.(\ref{eq:J}), is $\tau\sim 5\times
10^{-10}{\rm sec}.$ The ``black hole epoch" thus ended in time
before baryogenesis and other subsequent epochs in the standard
cosmology. As suggested in Ref.~\cite{bellido}, such a
post-inflation PBH evaporation provides an interesting mechanism
for reheating. Note that due to the continuous evaporation process
and the Hubble expansion, the BH reheating should result in an
effective temperature which is sufficiently lower than the Planck
scale.

\section{Discussion}

Our arguments for the existence of BHR based on GUP is heuristic.
The search for its deeper theoretical foundation is currently
underway\cite{CDS}. As interactions with BHR are purely
gravitational, the cross section is extremely small, and direct
observation of BHR seems unlikely. One possible indirect signature
may be associated with the cosmic gravitational wave background.
Unlike photons, the gravitons radiated during evaporation would be
instantly frozen. Since, according to our notion, the BH
evaporation would terminate when it reduces to a BHR, the graviton
spectrum should have a cutoff at Planck mass. Such a cutoff would
have by now been redshifted to $\sim {\mathcal O}(10^4)$ GeV.
Another possible GW-related signature may be the GWs released
during the gravitational collapse at $t\sim t_h$. The frequencies
of such GWs would by now be in the range of $\sim 10^7-10^8$ Hz.
It would be interesting to investigate whether these signals are
in principle observable. Another possible signature may be some
imprints on the CMB fluctuations due to the thermodynamics of
PBH-CMB interactions. These will be further investigated.

\section*{Acknowledgments}

I deeply appreciate my early collaborations and fruitful
discussions with Ronald J. Adler. I also thank K. Dasgupta, A.
Linde, M. Rees, M. Shmakova, J. Silk, P. Steinhardt, and M.
Tegmark for helpful discussions. This work is supported by the
Department of Energy under Contract No. DE-AC03-76SF00515.


\end{document}